**Controlling thermal cycling effect in phase separated manganites with high temperature thermal treatments.**

M. Quintero[1,2], B. Sievers[1,2], J. Sacanell[1,2].

[1]Departamento de Física de la Materia Condensada, Gerencia de Investigación y Aplicaciones, Centro Atómico Constituyentes, CNEA.

[2]Instituto de Nanociencia y Nanotecnología, CNEA-CONICET

Several phase separated manganites present a peculiar effect each time they go through a phase transition within the range characterized by phase separation. The effect is known as the thermal cycling effect (TCE) and is characterized by monotonous changes in the relative content of the coexisting phases.

In this work, we analyze a way to control the effects induced by TCE, performing thermal treatments at high temperature. Our results revealed a complex interplay between the dynamic and static characteristics of the phase separated state, which can be analyzed in terms of three simple parameters. One related to the static properties, another to the dynamic properties and a last one that acts as a link between both features.

**Introduction.**

The phase separation (PS) phenomenon, is the coexistence of different magnetic, electric and structural ordering observed in a large number of systems. It has significant influence on the physical properties of different compounds, in particular, in manganese oxides with mixed valence, also known as manganites [1].

The strong coupling between the different degrees of freedom of manganites (magnetic, electric and structural) provides a particular situation in which the coexisting phases evolve by competing one with each other, giving rise to a dynamic evolution of their physical properties [2,3,4,5,6,7,8,9]. Additionally, the relation between the coexisting phases can be unbalanced favoring one of them by changing the synthesis parameters [10,11,12] or by the application of external stimuli, such as electric [13,14,15,16] or magnetic fields [2,3,17]. This characteristic is particularly attractive for the development of technological devices in areas as for example spintronics [18], resistive RAM [19,20,21], etc.

A much less studied phenomenon, is the instability induced by thermal cycles or thermal cycling effect (TCE) observed in several manganites [22,23,24,25]. Changes in the relative content of the coexisting phases can be triggered by thermal cycling the system through the temperature range corresponding to phase separation.

The TCE has been studied in $La_{0.5}Ca_{0.5}MnO_3$ (LCMO), a prototypical manganite exhibiting PS between a metallic ferromagnetic (FM) and an insulating antiferromagnetic-charge ordered (AFM-CO) phases [11]. In this compound, a monotonous reduction of the FM/AFM fraction is the signature of the TCE, each time the system is subjected to a low temperature thermal cycle.

Within this framework the TCE phenomena and the dynamic evolution of the coexisting phases appear to be related as revealed the correlation between the blocking temperature and the relative FM fraction in the PS regime [26]. We have developed a model which accounts for two apparently unrelated phenomena: the TCE and the dynamical evolution of the amount of the coexisting phases if the system is kept at fixed temperature in the PS range. Assisted by TCE, the system evolves from cycle to cycle through the interface of the coexisting phases, retaining its memory of this thermal history in the subsequent measurement. Thus, the system can be trained by the accommodation strain between the FM and AFM-CO phases towards the most stable low temperature state, by a progressive release of structural strain [26]. Also, we established that the TCE can be suppressed by the application of an external magnetic field and hydrostatic pressure [25].

As with any change in the physical properties induced by an external feature it is possible to use the TCE in a technological device. For example, the TCE can be used to monitor the number of thermal cycles to which a particular device is subjected [27]. However, a study of the possibility to erase the changes induced by the TCE is lacking.

In this work we will focus our attention in the stability of the TCE when the sample is exposed to high temperature thermal treatments looking for the temperature that reverse the changes in the physical properties of LCMO induced by the TCE. An analysis in terms of the thermal history of the system is presented in order to contribute to the basic knowledge of the phenomenon.

**Experimental.**

Polycrystalline samples of LCMO were synthesized by the liquid mix technique, using 99.9% purity reactants. We followed the route detailed in reference [11], to obtain a sample with an average grain size of 450 nm, as estimated through SEM microphotographs. Chemical composition and crystalline structure were verified by EDS and X ray diffraction, respectively. Magnetization measurements were performed in a vibrating sample magnetometer Versalab$^{TM}$ manufactured by Quantum Design.

**Results and discussion**

In figure 1, we present the magnetization of the LCMO sample as function of temperature for one thermal cycle between 300 K and 50 K at 2 K/min with an applied magnetic field of 0.1T [28]. The system is paramagnetic (PM) at room temperature. On cooling the sample, a PM to FM transition can be observed at $T_C \sim 220$ K. A steep decrease of the magnetization indicates the formation of the AFM-CO phase into the FM matrix at $T_{co}$ below 150K, and on further cooling, the magnetization reaches a stable value which remains constant until the lowest measured temperature (50 K). If we assume that the magnetic signal is mainly due to the FM phase, then $M_{50K}$ is proportional to the low temperature relative FM phase fraction. The state of the sample in the alluded temperature range, is the consequence of the existence of a frozen PS regime, in which the relative fractions of the coexisting FM and CO phases remains blocked [5] until the sample reaches the lowest temperature of 50K. On warming, a dynamic PS regime appears, which is separated from the frozen PS state by a blocking temperature $T_B$, indicated by an arrow in figure 1.

A large thermal hysteresis is observed between the cooling and warming curves signaling the first order character of the FM to AFM-CO phase transition. Also, an instability has been observed against thermal cycling (i.e., the TCE), related with the competition between the coexisting phases and characterized by a cumulative monotonous reduction of the low temperature magnetization, from one 300K-50K cycle to another (inset of figure 1).

This last phenomenon has been observed in LCMO and in other manganites exhibiting PS [22,23,24,25] and manifests as irreversible changes in the physical properties in the range characterized by PS.

Previously [26], evidence has been presented that suggests that the strain between the FM and CO phases in coexistence produce defects that act as pinning centers for the FM/CO interface. In this picture, the TCE is related to a progressive release of strain from one thermal cycle (between 300 and 50 K) to the other. That release, brings the system closer to the most stable low temperature state, which in the case of LCMO is towards a reduction of the relative FM phase fraction. Thus, by the TCE, the system keeps memory of the state reached in previous cycles, a fact evidenced as a cumulative reduction in the low temperature magnetization [26] as we present in the inset of figure 1. This particular behavior is one of the most remarkable characteristics and the signature of the TCE.

Within this scenario, considering that the defects formed in the FM/CO interface in each thermal cycle are the responsible for the TCE, we can study the possibility to erase or revert the changes induced by the effect. In order to do that, we performed thermal treatments at temperatures above room T, but keeping the temperature of the thermal treatments below the one used in the synthesis procedure (~ 900ºC).

We performed sets of thermal cycles between 300 K and 50 K. Between each set we performed a high temperature thermal treatment (HTTT) where the sample temperature was kept constant at a fixed temperature $T_{HTTT}$ for three hours. In figure 2 we present the first thermal cycle performed after each HTTT. One of the most noticeable effects is the reduction of the thermal hysteresis for curves corresponding to $T_{HTTT} > 550$ K. Another interesting feature is the increase in the low temperature magnetization ($M_{50K}$) being indicative of the relative reduction of the fraction of CO phase.

In figure 3 we present the evolution of $M_{50K}$ during the whole sequence as a function of the cycle number. The HTTT are indicated with arrows and changing the color of the symbol. The TCE effect in LCMO is characterized by a progressive reduction of the magnetization in the PS temperature range, due to a reduction of the relative FM fraction. We can see that the application of HTTT at T > 400K provokes some kind of "inverse" effect, in the sense that the value of $M_{50K}$ increases after each HTTT is performed. However, the effect of HTTT is not a simple reversion of the effect, taking sample to its state a few cycles ago.

The increase of $M_{50K}$ after a HTTT is larger, the higher $T_{HTTT}$ is. We also see that the initial value of $M_{50K}$ for a particular series performed after an HTTT, can be larger than that of the initial value of the whole sequence. Additionally, after a treatment at 650 K, the subsequent reduction ascribed to the next TCE is somehow inhibited. A further HTTT at 675 K, provokes changes in the subsequent magnetization curves that suggest re-crystallization of the samples and thus, we are going to leave those measurements out of the present analysis.

For HTTT below 650 K, the evolution of $M_{50K}$ for each $T_{HTTT}$ can be described by [26],

$$M_{50K} = Ae^{-\frac{n}{\lambda}} + M_{EQ} \qquad (1)$$

with n the number of cycles starting after each HTTT, λ representing the constant of decay and $M_{EQ}$ the asymptotic equilibrium magnetization at 50K. In figure 4 we present λ and $M_{EQ}$ as function of $T_{HTTT}$. For $T_{HTTT} < 550K$, the value of $M_{EQ}$ remains almost constant, increasing significantly above this temperature.

A clear monotonous reduction of $\lambda$ for $T_{HTTT} > 400K$, pointing to a zero value at $T_{HTTT}$ close to 650 K. It is worth to note that at $T_{HTTT} = 625K$ the value of $M_{EQ}$ increases very significantly, as compared to its value below 550K. We cannot be sure if a peak is reached at this point because for $T_{HTTT} > 625K$, the TCE is lost and thus, the relevance of equation (1).

Summarizing, with the HTTT, we are changing the characteristic scale of the evolution given by the TCE (a smaller $\lambda$ means that a small number of cycles is needed for a particular reduction of the magnetization to be reached). The reduction of $\lambda$ from $T_{HTTT} = 400K$, until $T_{HTTT} > 625K$, point at which it reaches zero (figure 4), means that the change induced by the TCE is "faster" in a scale measured in cycles, as the temperature of the HTTT grows. In this sense, $\lambda = 0$ at $T_{HTTT} = 625K$ corresponds to the case in which no cycle is needed to reach equilibrium. This is consistent with the fact that for the HTTTs at $T_{HTTT} > 625K$, the TCE disappears. It is worth to note that this is not the only effect produced by the HTTT, as it also affects the particular equilibrium state, a fact that can be seen in the increasing values for $M_{EQ}$ after each treatment is performed. The fact that the TCE affects not only the static properties of a PS manganites as LCMO but also its dynamic behavior, has been already stablished [26]. The most significant feature regarding dynamic behavior, is the change in the blocking temperature of the system, which separates two PS regimes, a "dynamic PS", characterized by a logarithmic evolution of the relative content of the FM fraction and a "frozen PS", in which the evolution is blocked. The particular interplay between static and dynamic properties was evidenced in a linear correlation among $M_{50K}$ (static) and the inverse of $T_B$ (dynamic) [26]. This is consistent with the fact that $T_B$ is inversely proportional to the

*distance-to-equilibrium* of the system, characterized by the difference between the equilibrium and the actual FM relative fractions [6,29].

With the new results, a correlation among $T_B$ and $\lambda$ is obtained, that separates two regimes as a function of $T_{HTTT}$. In figure 5 we present the correlation among $\lambda$ and $1/(T_B-T_{B0})$, being $T_{B0}$ the blocking temperature observed before the application of any HTTT. The observed linear behavior is indicative of the strong coupling between both quantities.

The previously reported correlation between $M_{50K}$ and the inverse of $T_B$ along with the correlation between $\lambda$ and the inverse of $T_B$ can be interpreted in terms of the significance of $\lambda$. This parameter is the "rate of decay" from the actual value of $M_{50K}$ reached in a particular case, which is in a blocked PS state, to the $M_{EQ}$ corresponding to the equilibrium condition at that situation.

As the HTTT provokes a change in $M_{EQ}$ and thus in $T_B$, this change is also reflected in $\lambda$, which has been shown to be rate dependent [26]. Thus, $\lambda$ acts as a link between the static and dynamic properties, reflected by $M_{EQ}$ and $T_B$, respectively. This fact can be shown here by the coordinate change of the three reported parameters.

**Conclusions**

In this work we presented a study related to the thermal cycling effect observed in $La_{0.5}Ca_{0.5}MnO_3$, aimed to explore the possibility to erase or revert the effect. We performed high temperature thermal treatments with temperatures above 300K and obtained a more complex scenario.

We observed several changes in the behavior of the magnetization curves: progressive reduction of the thermal hysteresis, modification in the low temperature magnetization and variation in the parameter that quantifies the rate of change ($\lambda$) for the TCE. Additionally, the blocking temperature is also modified by the HTTT. The close relation between this parameter and the dynamic behavior has been showed.

Several features appear, related with the TCE, that can be affected by the HTTT. The gradual inhibition of the first order character and the change in the low temperature magnetization (which is the starting point for the value of $M_{50K}$ in a set of cycles, after each HTTT), a parameter related with the static properties of the PS state. Along with the increase of the starting point of $M_{50K}$ after each HTTT, an increase in $M_{EQ}$ (its asymptotic value after several cycles) is also observed, indicating that the HTTT role is not as simple as a reversion of previously occurred TCE related changes. The actual equilibrium state (given by $M_{EQ}$) is also affected.

The peak of $M_{EQ}$, correlates with the tendency to zero of $\lambda$, showing a clear correlation between the static and dynamic properties of the PS state of LCMO. This correlation has already been established by studying the low temperature properties of LCMO, but here, the additional correlation between $\lambda$ and $T_B$, positions $\lambda$ as a link among the dynamic and static properties that combines features of both aspects of the PS state.

A deep analysis has to be performed on the dependence of the rate of change constant, $\lambda$, as a function of the cooling heating rate, magnetic field to further understand and complement the presented results.

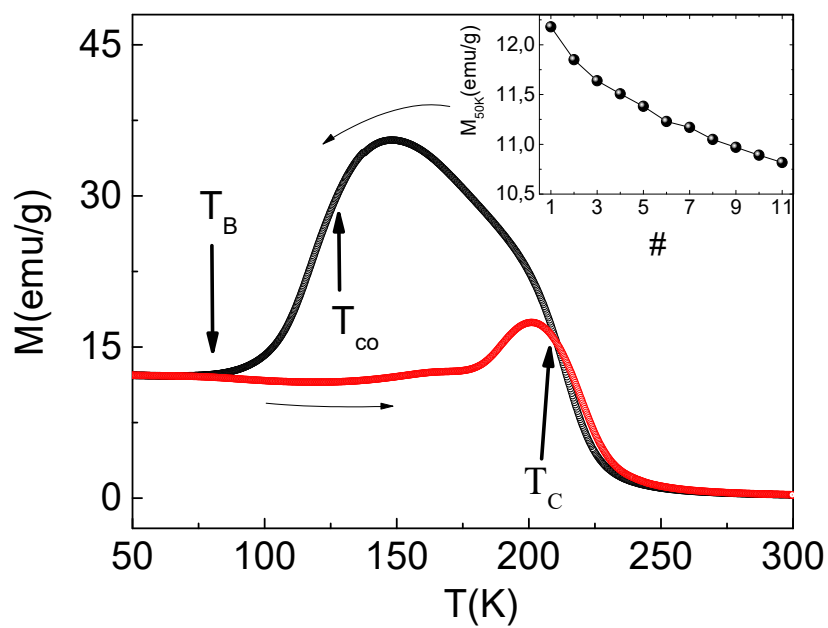

Figure 1: Magnetization as function of temperature (H = 0.1T). Inset: Magnetization at 50K ($M_{50K}$) as a function of the number of the thermal cycle.

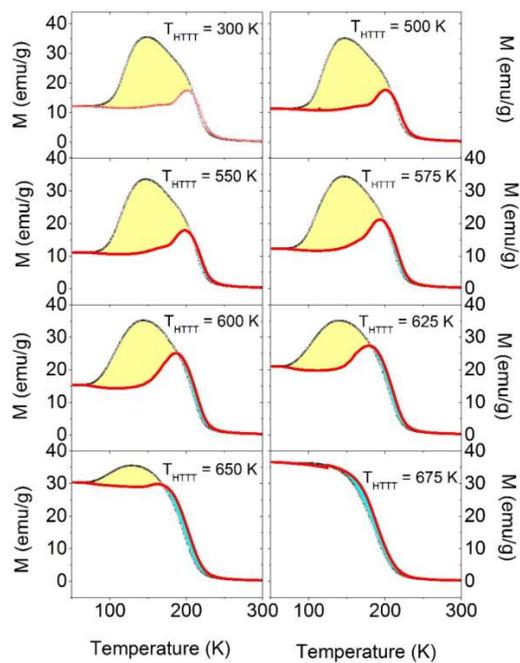

Figure 2: Magnetization as function of temperature after different high temperature thermal treatment (HTTT).

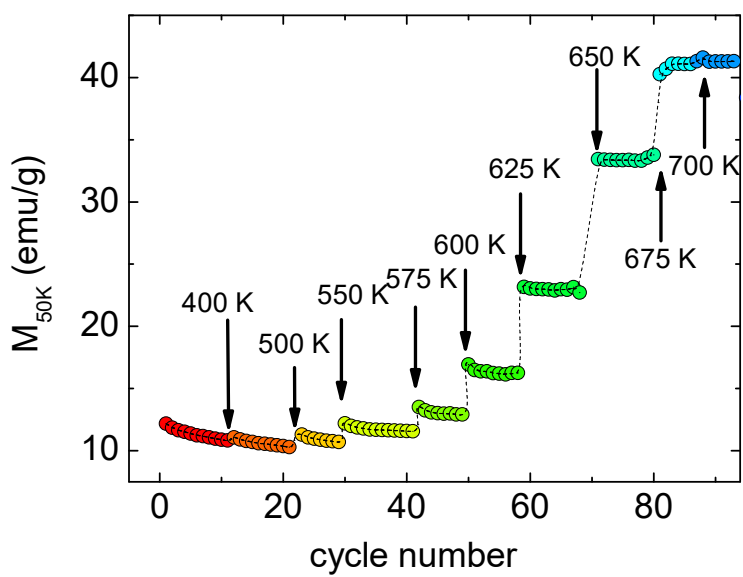

Figure 3: Magnetization at 50K as function of the cycle number in the complete sequence of the sample´s thermal history, including the high temperature thermal treatments (HTTT), marked with arrows.

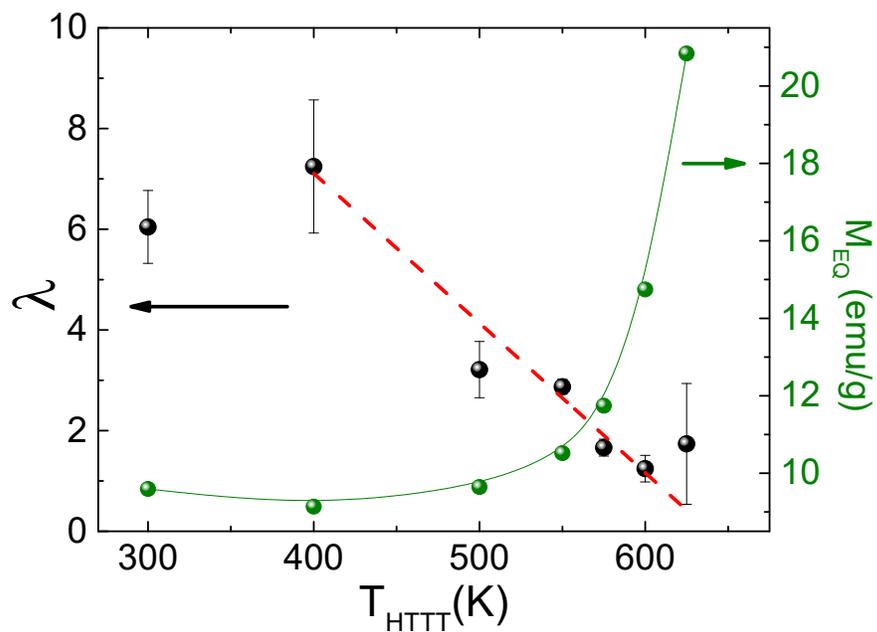

Figure 4: Decay constant for the TCE (λ) and asymptotic value for the magnetization at 50K ($M_{EQ}$) according to eq. 1, as function of the temperature of the HTTT ($T_{HTTT}$).

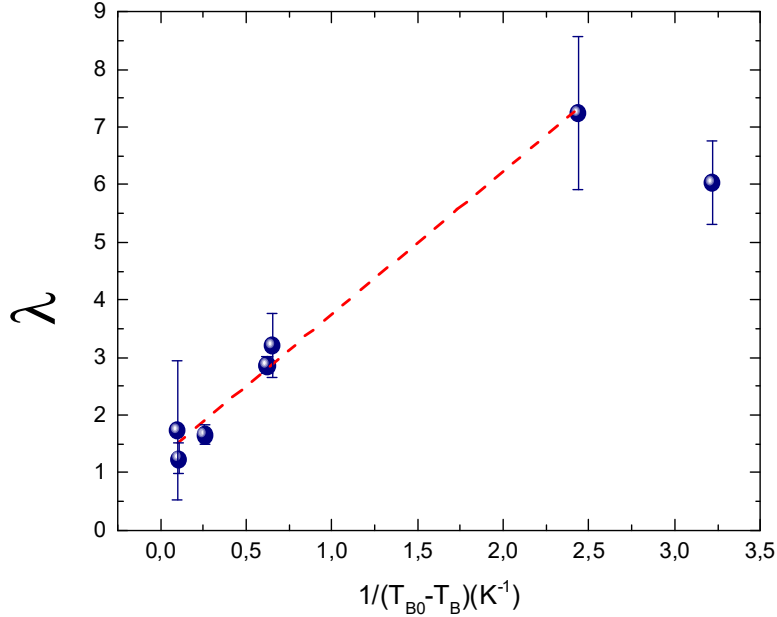

Figure 5: Decay constant for the TCE ($\lambda$) vs $1/(T_B-T_{B0})$.